\documentclass[prl,twocolumn,showpacs,preprintnumbers,amsmath,amssymb]{revtex4}

\usepackage{graphicx}
\usepackage{dcolumn}
\usepackage{bm}

\begin{document}

\title{Temperature of the inner-core boundary of the Earth: Melting of iron at 
high pressure from first-principles coexistence simulations }

\author{Dario Alf\`{e}}\email{d.alfe@ucl.ac.uk}
\affiliation{
  Materials Simulation Laboratory, and London Centre for Nanotechnology \\
  Department of Earth Sciences, and Department of Physics and Astronomy, \\
  UCL, Gower Street, London WC1E 6BT, United Kingdom  }

\date{\today}

\begin{abstract}
  The Earth's core consists of a solid ball with a radius of 1221 Km,
  surrounded by a liquid shell which extends up to 3480 Km from the
  centre of the planet, roughly half way towards the surface (the mean
  radius of the Earth is 6373 km). The main constituent of the core is
  iron, and therefore the melting temperature of iron at the pressure
  encountered at the boundary between the solid and the liquid (the
  ICB) provides an estimate of the temperature of the core. Here I
  report the melting temperature of Fe at pressures near that of the
  ICB, obtained with first principles techniques based on density
  functional theory. The calculations have been performed by directly
  simulating solid and liquid iron in coexistence, and show that and
  at a pressure of $\sim 328$~GPa iron melts at $\sim 6370\pm
  100$~K. These findings are in good agreement with earlier
  simulations, which used exactly the same quantum mechanics
  techniques, but obtained melting properties from the calculation of
  the free energies of solid and liquid
  Fe~\cite{alfe1999,alfe2002,alfe2001}.
\end{abstract}

\maketitle

The study of iron under extreme conditions has a long hystory. In
particular, numerous attempts have been made to obtain its high
pressure melting properties~\cite{williams87, boehler93, shen98,
  jephcoat96, saxena94, ma04, brown86, nguyen04,
  yoo93}. Experimentally, Earth's core conditions can only be
reproduced by shock wave (SW) experiments, in which a high speed
projectile is fired at an iron sample, and upon impact high pressure
and high temperature conditions are produced. By varying the speed of
the projectile it is possible to investigate a characteristic
pressure-volume relation known as the Hugoniot~\cite{poirier91}, and
even infer temperatures, although a word of caution here is in
order, as temperature estimates are often based on the knowledge of
quantities like the constant volume specific heat and the Gr\"uneisen
parameter, which are only approximately known at the relevant
conditions~\cite{brown86}. If the speed of the projectile is high
enough, the conditions of pressure and temperature are such that the
sample melts, and it is therefore possible to obtain points on the
melting curve, of course with the caveat mentioned above about
temperature measurements. An alternative route to high pressure high
temperature properties is the use of diamond anvil cells (DAC), in
which the sample is surrounded by a pressure medium and statically
compressed between two diamond anvils. In DAC experiments pressure and
temperatures can be directly measured, and therefore these techniques
should in principle be more reliable to investigate melting
proeprties. Unfortunately, in the case of iron is it not so, and there
is a fairly large range of results obtained by different
groups~\cite{williams87, boehler93, shen98, jephcoat96, saxena94,
  ma04}.

An alternative approach used for the past ten years or so has been to
employ theory --and in particular quantum mechanics techniques based on
density functional theory-- to calculate the high pressure melting
curve of iron. A number of groups have used different approaches to
the problem. Our own strategy has been to calculate the Gibbs free
energy of solid and liquid iron, and then obtain the melting curve by
imposing their equality for any fixed pressure. We obtained a melting
temperature of $\sim 6350$~K at 330 GPa~\cite{alfe2002}. The approach
of Belonoshko et al.~\cite{belonoshko2000} was to fit an embedded atom
model (EAM) to first principles calculations, and then calculate the
melting curve of the EAM. They obtained a temperature of $\sim 7050$~K
at 330 GPa. The approach of Laio et al.~\cite{laio2000} was similar,
altough they refitted their optimised model potential (OPM) to first
principles calculation in a self-consistent way. They obtained a
melting temperature of $\sim 5400$~K at 330 GPa. We later re-conciled
the results of Belonoshko et al~\cite{belonoshko2000} with ours, by
showing that the difference was due to a difference in free energies
between their EAM and our DFT~\cite{alfe2002b}. A similar argument
would be responsible for the difference between our results and those
of Laio et al~\cite{laio2000}.

Here I am using an approach to melting which is independent from the
free energy technique used earlier~\cite{alfe1999, alfe2001,
  alfe2002}, and the main motivation of this work is to provide an
alternative route to the calculation of the melting properties of
Fe. The method employed here is that of the coexistence of phases, in
which solid and liquid iron are simulated in coexistence. The first
time that the method was used in the context of first principles
calculations was for the low pressure melting curve of
aluminium~\cite{alfe2003}, where it was shown to deliver the same
results as the free energy method~\cite{vocadlo2002}. It was later
applied to compute the melting curve of LiH~\cite{ogitsu03},
hydrogen~\cite{bonev04} and MgO~\cite{alfe2005}.

The coexistence method is intrinsically expensive, as it requires
large simulation cells and long simulations. It can be applied in a
number of diffent ways, here I have used the $NVE$ ensemble,
i.e. constant number of atoms $N$, constant volume $V$ and constant
internal energy $E$. In the $NVE$ ensemble, for each chosen volume $V$
there is a whole range of energies $E$ for which solid and liquid can
coexist for long time; the average temperature and pressure along the
simulation then provide a point on the melting curve. If the energy
$E$ is above(below) the range for which coexistence can be maintained,
the system will completely melt (solidify), and the simulation does
not provide useful melting properties informations. It should be
pointed out that any finite system will eventually melt or solidify if
simulated for long enough, due to spontaneous fluctuations. However,
melting(solidification) resulting from a too high(low) value of $E$
typically appear on much shorter time scales.

The present calculations have beed performed with density functional
theory with the generalised gradient approximation known as
PW91~\cite{pw91} and the projector augmented wave
method~\cite{blochl94,kresse99} as implemented in the {\sc VASP}
code~\cite{vasp}. An efficient extrapolation of the charge density was
employed~\cite{alfe1999b}. Single particle orbitals were expanded in
plane waves with a cutoff of 300 eV, and I used the finite temperature
implementation of DFT as developed by Mermin~\cite{mermin1965}. These
settings are exactly equivalent to those used in our previous
work~\cite{alfe1999,alfe2001,alfe2002}, so the melting properties
obtained here will be directly comparable to those early ones.  The
simulations have been performed on hexagonal closed packed (hcp) cells
containing 980 atoms ($7\times 7 \times 10$), using the $\Gamma$ point
only. For the temperatures of interest here the use of the $\Gamma$
point provides completely converged results.  The time step in the
molecular dynamics simulations was 1 femto-second (fs), and the
self-consistency on the total energy $ 2 \times 10^{-5}$~eV. With
these prescriptions the drift in the constant of motion was $\sim
0.5$~K/pico-second.

The coexistence simulations were prepared by starting from a perfect
hcp crystal, which was initially thermalised to $\sim 6300$~K for 1
pico-second (ps). Then half of the atoms in the cell were clamped and
the temperature was raised to a very high value, to melt the other
half of the cell. Once a good melt was obtained, the temperature was
reduced back to 6300~K and the system thermalised for one additional
ps, after which the simulation was stopped, new initial velocities
were assigned to the atoms and the simulation continued in the
micro-canonical ensemble. The simulations were monitored using the
density profile, calculated by dividing the simulation cell in 100
slices parallel to the solid-liquid interface and counting the number
of atoms in each slice; in the solid region this is a periodic
function, with large number of atoms if the slice coincide with an
atomic plane, and small values if it falls between atomic layers. In
the liquid region it fluctuates randomly around some average value.

I performed five different simulations, starting with different
amounts of internal energies $E$, provided to the system by assigning
different initial velocities to the atoms. The simulation with the
highest value of $E$ completely melted after $\sim 6 $~ps. The one
with the lowest amount of $E$ solidified after $\sim 11 $~ps. Among
the other three, one melted after $\sim 14$~ps, one after $\sim
24$~ps, while the last one has remained in coexistence for the whole
length of $\sim 25$~ps. However, most of these simulations were
coexisting for long enough, so that useful melting information from
the period of coexistence could actually be extracted in almost all
cases.

\begin{figure}
\centerline{
\includegraphics[width=3.4in]{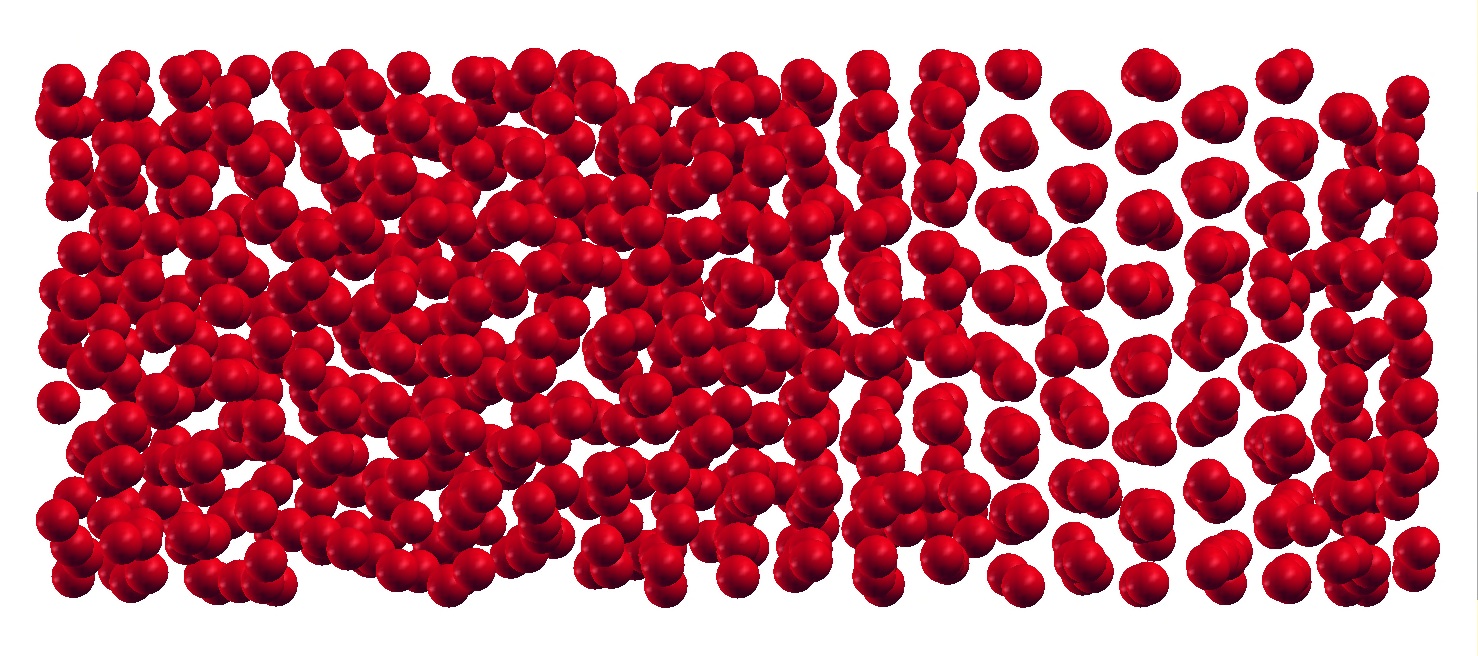}
}
\caption{(Color online) Snapshot of a DFT molecular dynamics
  simulation showing solid and liquid iron in coexistence. The
  simulation cell contains 980 atoms.}\label{fig:coexist}
\end{figure}

A snapshot of a simulation with solid and liquid in coexistence is
show in Fig.~\ref{fig:coexist}~\cite{xcrysdens}.

\begin{figure}
\centerline{
\includegraphics[width=3.4in]{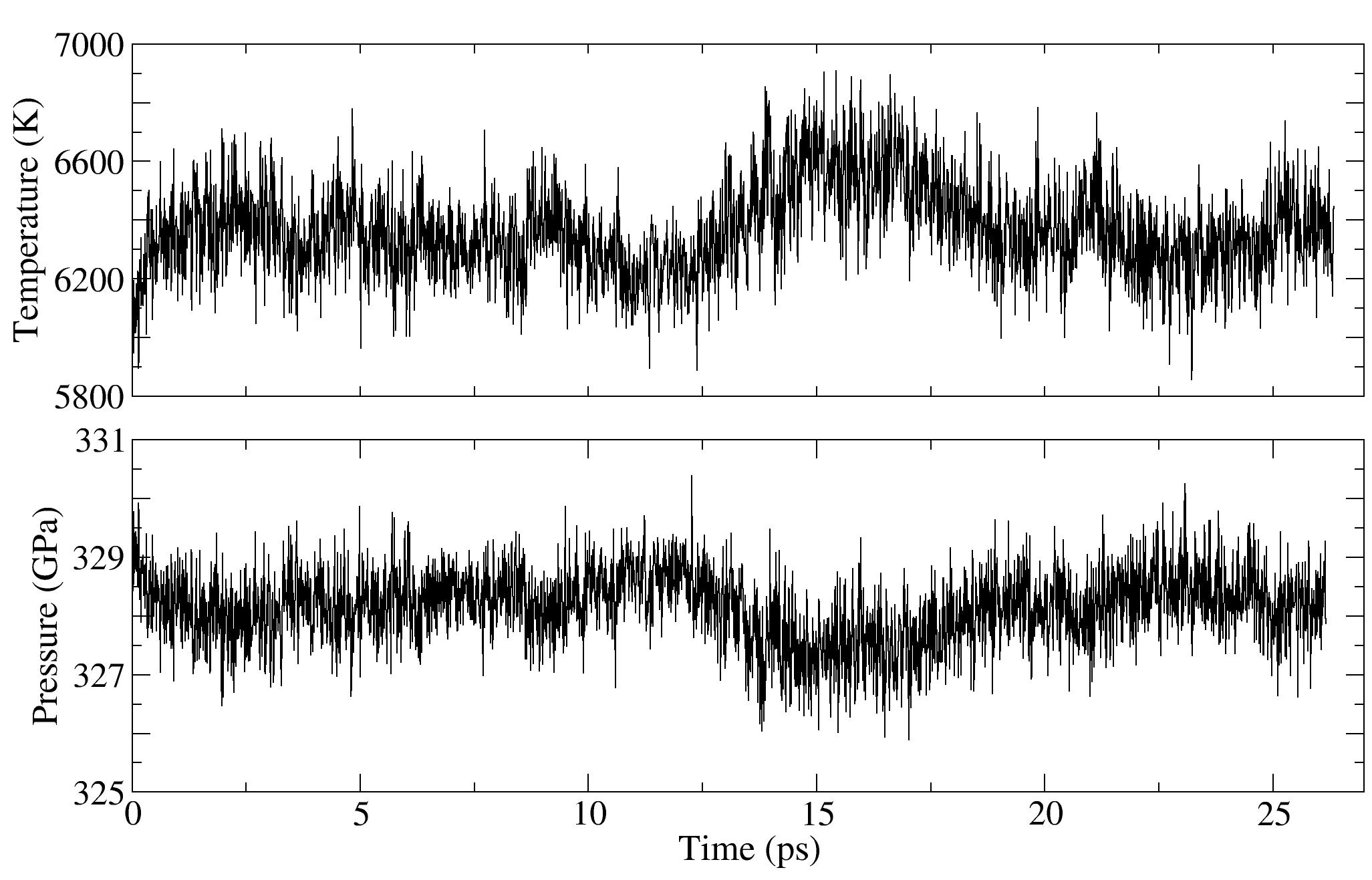}
}
\caption{Temperature (upper panel) and pressure (lower panel) for a
  simulation of solid and liquid iron in coexistence.}\label{fig:pt}
\end{figure}

In Fig.~\ref{fig:pt} I display the temperatures and the pressures
corresponding to the simulation that remained in coexistence for the
whole 25 ps length, which provides a melting point $(p,T) = (328 \pm
1$~GPa$, 6370 \pm 100$~K$)$. It is interesting to notice a temperature
excursion in the simulation after $\sim 15$~ps, which lasts for $\sim
5$~ps. This temperature variation is anti-correlated to a pressure
variation, and corresponds to a temporary loss of some liquid in the
cell, with latent heat of fusion converted into kinetic energy, and
volume of fusion responsible for the drop in pressure. Large
excursions of these type may provoke {\em accidental} melting (or
freezing), even if the internal energy $E$ is within the range of
coexistence. This problem is mitigated by the use of large simulation
cells, and therefore this is one of the reasons why large simulation
cells are needed in conjunction with the coexistence approach.

\begin{figure}
\centerline{
\includegraphics[width=3.4in]{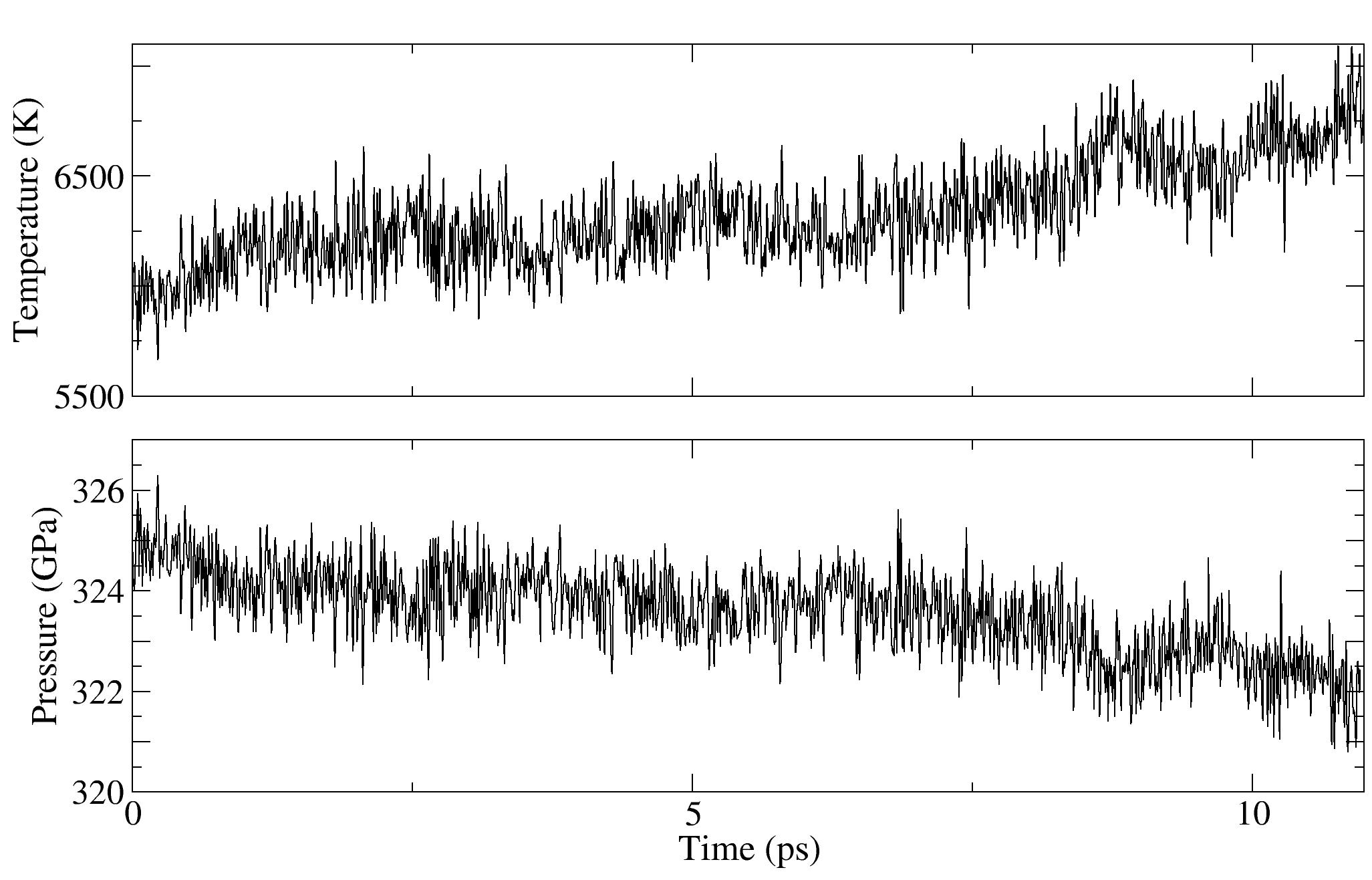}
}
\caption{Temperature (upper panel) and pressure (lower panel) for a
  simulation of solid and liquid iron in coexistence. The system
  eventually completely solidifies, with a drop in pressure and an
  increase in temperature due to release of volume and latent heat of
  fusion respectively.}\label{fig:pt2}
\end{figure}

In Fig.~\ref{fig:pt2} I show a simulation that eventually solidified,
however, as mentioned above, coexistence was maintained for a long
period, and the information gathered by the central part of the
simulation can still be used to obtain a point on the melting curve,
and the result is $(p,T) = (324 \pm 1$~GPa$, 6250 \pm 100$~K$)$, which
is consistent with the previous point.

\begin{figure}
\centerline{
\includegraphics[width=3.4in]{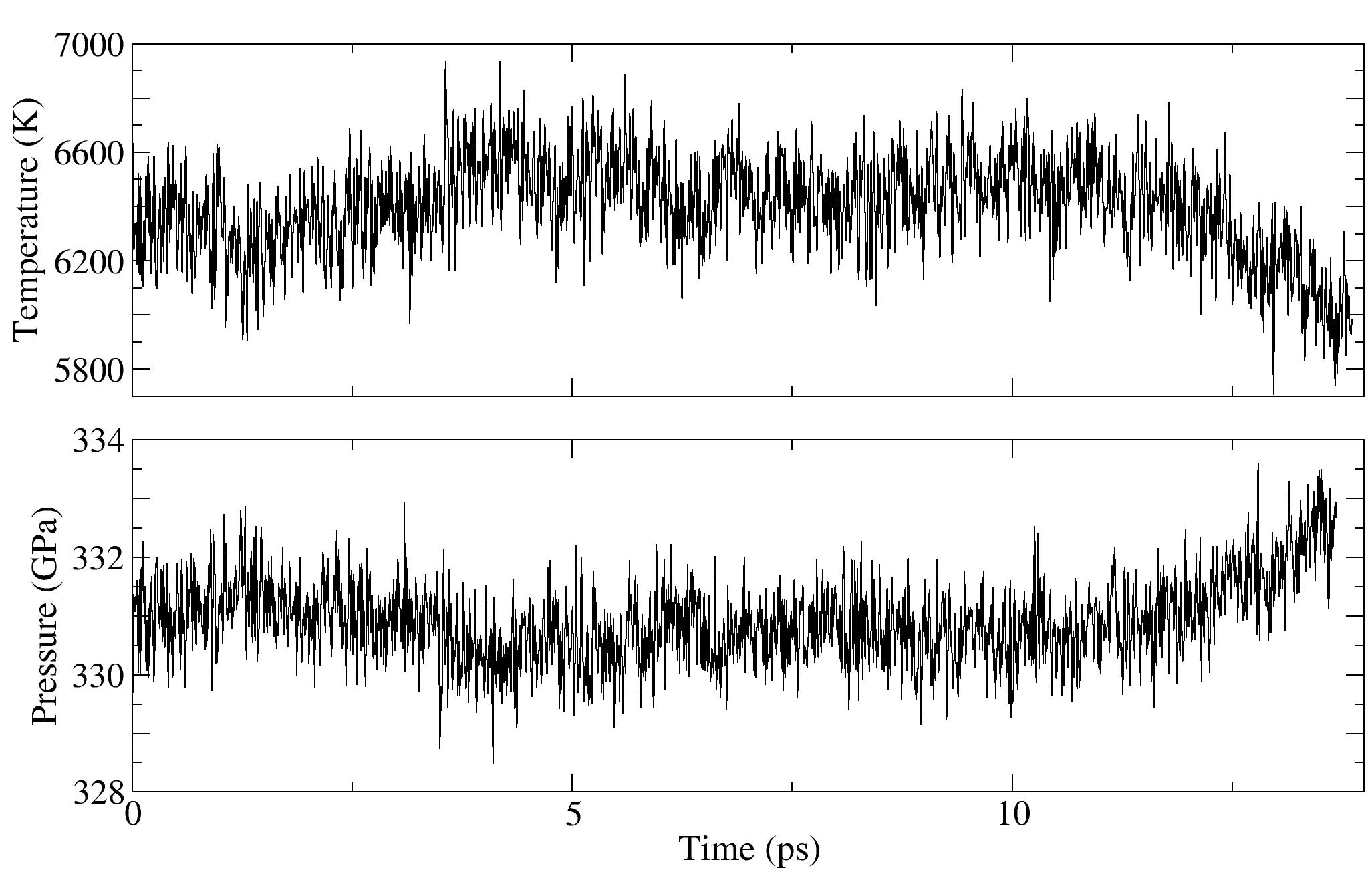}
}
\caption{As in Fig.~\ref{fig:pt2}, but here the system eventually
  melts.}\label{fig:pt3}
\end{figure}

Similarly, in Fig.~\ref{fig:pt3} I show one of the simulations that
eventually melted, and by taking the average temperature and pressure
from the central part of the simulation I get $(p,T) = (331 \pm
1$~GPa$, 6430 \pm 100$~K$)$, which is also consistent with the other
previous two points.

The $c/a$ value used in the simulations was fixed at 1.6, which
resulted in slightly non-hydrostatic conditions. To study the effect
of non-hydrostaticity, and that of the relatively small size of the
simulation cells, I have performed simulations using a classical
embedded atom model (EAM), adapted to deliver results very close to
the present ab-inition techniques~\cite{alfe2002b}. I have performed
simulations on cells containing 7840 atoms ($14 \times 14 \times 20$),
and found that at a pressure of 324 GPa the effect of using a small
cell containing only 980 atoms is to raise the melting temperature by
$\sim 100$~K. The ab-initio non-hydrostatic conditions with $c/a =
1.6$ are similarly reproduced by the EAM, which also shows that with
$c/a = 1.65$ the simulations are almost exactly under hydrostatic
conditions~\cite{hydrostatic}.  The effect of the non-hydrostaticity
with $c/a=1.6$ is to reduce the melting temperature by $\sim 100$~K,
so that the combined effects of non-hydrostaticity and small size
cancel each other. A final check was performed by repeating the
simulations using 62720 atom-cells ($28 \times 28 \times 40$), which
showed essentially no differences with the results obtained using 7840
atom-cells.

\begin{figure}
\centerline{
\includegraphics[width=3.4in]{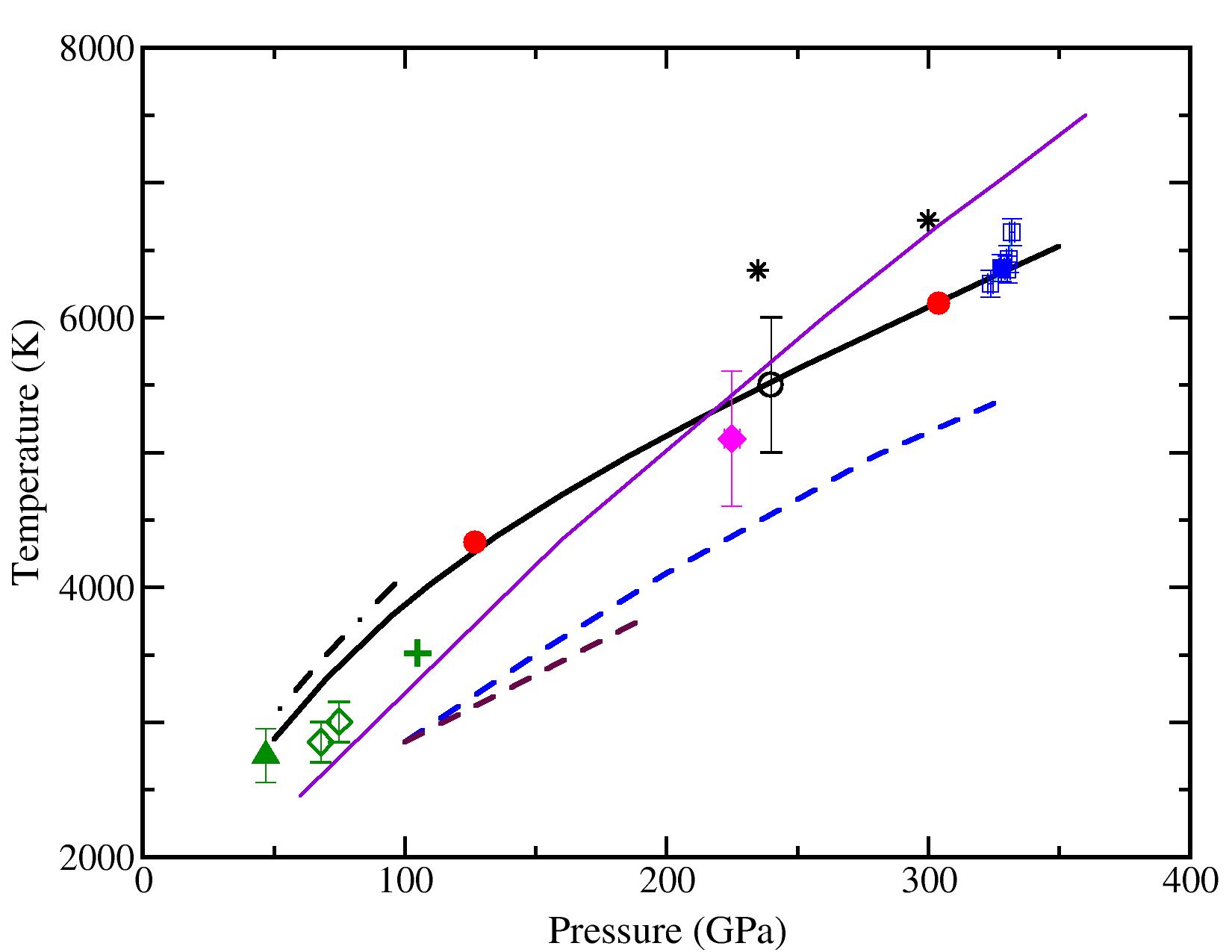}
}
\caption{(Color online) Comparison of melting curve of Fe from present
  calculations with previous experimental and {\em ab initio} results;
  blue filled square: DFT coexisting simulation for over 25 ps; blue
  open squares: DFT coexisting simulations ending up to complete
  solids or liquids, results are gathered from the region of
  coexistence; black heavy solid curve: DFT melting curve from
  Ref.~\protect\cite{alfe2002}; long-dashed blue curve: DFT-OPM results
  of Ref.~\protect\cite{laio2000}; light purple solid curve:
  DFT-EAM results of Ref.~\protect\cite{belonoshko2000}; filled red
  circles: DFT-EAM ``corrected'' results (see text); black chained and
  maroon dashed curves: DAC measurements of
  Refs.~\protect\cite{williams87} and ~\protect\cite{boehler93}; green
  open diamonds: DAC measurements of Ref.~\protect\cite{shen98}; green
  plus: DAC measurement of Ref.~\protect\cite{ma04}; green filled
  triangle: DAC measurement of Ref.~\protect\cite{jephcoat96}; black
  stars, black open circle and pink filled diamond: shock experiments
  of Refs.~\protect\cite{yoo93},~\protect\cite{brown86}
  and~\protect\cite{nguyen04}. Error bars are those quoted in original
  references. }
\label{fig:melt}
\end{figure}

All the present first-principles coexistence results are displayed in
Fig.~\ref{fig:melt}, which also contains experimental and previous
theoretical results. The filled square corresponds to the simulation
which maintained coexistence throughout (Fig.~\ref{fig:pt}), while the
empty squares to the other 4 simulations, for which the final part has
been discarded. As noted above, it is clear that all simulations
provide similar melting points, with the exception perhaps of the
point correponding to the simulation that melted only after $\sim
6$~ps (point at highest temperature in Fig~\ref{fig:melt}).  The
comparison of the present results with the earlier melting curve
obtained using the free energy approach~\cite{alfe2002} shows
excellent agreement, and the two sets of data therefore support each
other. I also report on the same figure the DAC experiments of
Refs.~\cite{boehler93, shen98, williams87, jephcoat96, ma04}, the SW
experiments of Refs.~\cite{brown86, nguyen04, yoo93} and the
calculations of Refs.~\cite{laio2000, belonoshko2000}. As mentioned
above, we identified the reasons of the differences between our
melting curve and that calculated by Belonoshko et
al.~\cite{belonoshko2000} being the free energy differences between
their EAM and our DFT. Once these differences are taken into account
it is possible to ``correct'' the EAM melting curve. The two red dots
on the figure show the corrected EAM results at two different
pressures, which agree with our DFT melting curve.

In conclusion, I have presented here calculations for the melting
temperature of iron under Earth's core conditions, obtained with
density functional theory and the technique of the coexistence of
phases. The DFT techniques are the same as those employed in our
earlier calculations, where we computed the melting curve using the
free energy approach~\cite{alfe1999,alfe2001,alfe2002}. The
consistence between the present calculations and those early ones is
expected, because the DFT technicalities are the same, and provides an
independent check on the accuracy of those early free energy
calculations. The DFT melting temperature of Fe at a pressure of $\sim
330$~GPa is therefore confirmed to be in the region of $\sim
6300-6400$~K.

The next question that one might ask is how accurate is DFT for this
problem. We argued in our previous work that DFT should indeed be
quite reliable, mainly because solid and liquid iron have very similar
structures, so that possible errors would largely cancel between the
two phases. However, we also pointed out that DFT does not seem to
reproduce the zero temperature pressure-volume equation of state of
hcp iron completely correctly, possibly underestimating the pressure
by $\sim 2.5\%$. We then argued that this error could propagate to the
melting curve, resulting in a lowering of temperatures which at a
pressure of 330 GPa could be in the region of $\sim
150$~K~\cite{alfe2002}. This would bring the melting temperature of Fe
at ICB condition to $\sim 6200$~K.  It will be interesting to re-visit
this problem with more accurate quantum mechanics techniques, and we
are planning to do so by using quantum Monte Carlo. We will report on
these results in due course.

This work was conducted as part of a EURYI scheme award as provided by
EPSRC (see www.esf.org/euryi). Calculations have been performed on the
UK national facility HECToR, using allocation of time from the Mineral
Consortium and from a EPSRC Capability Challenge grant. Calculations
were also performed on the UCL research-computing facility Legion, and
initially on the Cambridge High Performance facility Darwin.
Simulations were typically run on 256 cores, each molecular dynamics
step of 1 fs taking $\sim 7.5$~minutes.


\begin{thebibliography}{99}

\bibitem{alfe1999} D. Alf\`e, M. J. Gillan and G. D. Price, Nature, {\bf
    401}, 462 (1999).

\bibitem{alfe2002} D. Alf\`e, G. D. Price, and M. J. Gillan, Phys.
  Rev. B, {\bf 65}, 165118, (2002).

\bibitem{alfe2001} D. Alf\`e, G. D. Price, M. J. Gillan, Phys. Rev. B,
{\bf 64}, 045123 (2001).

\bibitem{williams87} Q. Williams, R. Jeanloz, J. D. Bass,
B. Svendesen, T. J. Ahrens, Science {\bf 286}, 181 (1987).

\bibitem{boehler93}
R. Boehler, Nature {\bf 363}, 534 (1993).

\bibitem{shen98}
G. Shen, H. Mao, R. J. Hemley, T. S. Duffy and M. L. Rivers,
Geophys. Res. Lett. {\bf 25}, 373 (1998).

\bibitem{jephcoat96} A. P. Jephcoat and S. P. Besedin,
Phi. Trans. R. Soc. Lond. A {\bf 354}, 1333 (1996).

\bibitem{saxena94}
S. K. Saxena, G. Shen, and P. Lazor, Science, {\bf 264}, 405 (1994). 

\bibitem{ma04} Y. Ma, M. Somayazulu, G. Shen, H. K. Mao, J. Shu,
R. J. Hemley, Phys. Earth Planet. Int., {\bf 143-144}, 455 (2004).

\bibitem{brown86} J. M. Brown and R. G. McQueen, J. Geophys. Res. {\bf
91}, 7485 (1986).

\bibitem{nguyen04} J. H. Nguyen, and N. C. Holmes, Nature {\bf 427}
339 (2004).

\bibitem{yoo93}
C. S. Yoo, N. C. Holmes, M. Ross, D. J. Webb and C. Pike,
Phys. Rev. Lett. {\bf 70}, 3931 (1993).

\bibitem{poirier91}
J.-P. Poirier, {\em Introduction to the Physics of the Earth's
Interior}, Cambridge University Press, Cambridge (1991).

\bibitem{belonoshko2000} A. B. Belonoshko, R. Ahuja, and B. Johansson,
  Phys. Rev. Lett., {\bf 84}, 3638 (2000).

\bibitem{laio2000} A. Laio, S. Bernard, G. L. Chiarotti, S. Scandolo and
  E. Tosatti, Science {\bf 287}, 1027 (2000).

\bibitem{alfe2002b} D. Alf\`e, G. D. Price, M. J. Gillan, J.
Chem. Phys, {\bf 116}, 7127 (2002).

\bibitem{alfe2003} D. Alf\`e, Phys. Rev. B, {\bf 68}, 064423 (2003).

\bibitem{vocadlo2002} L. Vo\v{c}adlo and D. Alf\`e, Phys. Rev. B, {\bf
65}, 214105 (2002).

\bibitem{ogitsu03} T. Ogitsu, F. Schwegler, F. Gygi, G. Galli,
Phys. Rev. Lett., {\bf 91}, 175502 (2003)

\bibitem{bonev04} S. A. Bonev, F. Schwegler, T. Ogitsu, G. Galli,
Nature {\bf 431}, 669 (2004).

\bibitem{alfe2005} D. Alf\`e, Phys. Rev. Lett., {\bf 94}, 235701 (2005).

\bibitem{pw91} Y. Wang and J. Perdew, Phys. Rev. B {\bf 44}, 13298
(1991); J. P. Perdew, J. A. Chevary, S. H. Vosko, K. A. Jackson,
M. R. Pederson, D. J. Singh and C. Fiolhais, Phys. Rev. B {\bf 46},
6671 (1992).

\bibitem{vasp} G. Kresse and J. Furthm\"{u}ller, Phys. Rev. B {\bf
    54}, 11169 (1996).
\bibitem{blochl94} P. E. Bl\"{o}chl, Phys. Rev. B {\bf 50}, 17953
  (1994).
\bibitem{kresse99} G. Kresse and J. Joubert, Phys. Rev. B {\bf 59},
  1758 (1999).
\bibitem{pbe} J. P. Perdew, K. Burke and M. Ernzerhof,
  Phys. Rev. Lett. {\bf 77}, 3865 (1996).
\bibitem{alfe1999b} D. Alf\`e, Comp. Phys. Comm. {\bf 118}, 31 (1999).

\bibitem{mermin1965} N. D. Mermin, Phys. Rev. {\bf 137}, A1441 (1965).

\bibitem{xcrysdens} Figure realised with the xcrysdens software:
  A. Kokalj, Comp. Mater. Sci. {\bf 28}, 155 (2003). Code available
  from http://www.xcrysden.org/.

\bibitem{hydrostatic} Simulations were also performed in the $NpH$
  ensemble (constant pressure $p$ and enthalpy $H$), using the
  algorithm developed by E.R. Hernandez, J. Chem. Phys. {\bf 115},
  10282 (2001).

\end{thebibliography}
\end{document}